\documentclass[conference]{IEEEtran}
\IEEEoverridecommandlockouts
\usepackage{cite}
\usepackage{amsmath,amssymb,amsfonts}
\usepackage{algorithmic}
\usepackage{graphicx}
\usepackage{textcomp}
\usepackage{xcolor}
\usepackage{todonotes}
\usepackage{amsmath}
\usepackage{caption}
\usepackage{subcaption}
\usepackage{hyperref}
\usepackage{fancyhdr}

\def\BibTeX{{\rm B\kern-.05em{\sc i\kern-.025em b}\kern-.08em
    T\kern-.1667em\lower.7ex\hbox{E}\kern-.125emX}}
\begin{document}

\title{Label Uncertainty Modeling and Prediction for Speech Emotion Recognition using \emph{t}-Distributions\\
\thanks{This work was supported by the Landesforschungsförderung Hamburg (LFF-FV79), under the ''Mechanisms of Change in Dynamic Social Interaction'' project.}

}

  


\author{\IEEEauthorblockN{Navin Raj Prabhu}
\IEEEauthorblockA{\textit{Signal Processing} \\
\textit{Universit\"at Hamburg}\\
Hamburg, Germany \\
navin.raj.prabhu@uni-hamburg.de}
\and
\IEEEauthorblockN{Nale Lehmann-Willenbrock}
\IEEEauthorblockA{\textit{Industrial and Organizational Psychology} \\
\textit{Universit\"at Hamburg}\\
Hamburg, Germany \\
nale.lehmann-willenbrock@uni-hamburg.de}
\and
\IEEEauthorblockN{Timo Gerkmann}
\IEEEauthorblockA{\textit{Signal Processing} \\
\textit{Universit\"at Hamburg}\\
Hamburg, Germany \\
timo.gerkmann@uni-hamburg.de}
}


\maketitle

\thispagestyle{fancy}

\begin{abstract}
As different people perceive others' emotional expressions differently, their annotation in terms of arousal and valence are per se subjective. To address this, these emotion annotations are typically collected by multiple annotators and averaged across annotators in order to obtain labels for arousal and valence. However, besides the average, also the uncertainty of a label is of interest, and should also be modeled and predicted for automatic emotion recognition. In the literature, for simplicity, label uncertainty modeling is commonly approached with a Gaussian assumption on the collected annotations. However, as the number of annotators is typically rather small due to resource constraints, we argue that the Gaussian approach is a rather crude assumption. In contrast, in this work we propose to model the label distribution using a Student's $t$-distribution which allows us to account for the number of annotations available. With this model, we derive the corresponding Kullback-Leibler  divergence based loss function and use it to train an estimator for the distribution of emotion labels, from which the mean and uncertainty can be inferred. Through qualitative and quantitative analysis, we show the benefits of the $t$-distribution over a Gaussian distribution. We validate our proposed method on the AVEC'16 dataset. Results reveal that our $t$-distribution based approach improves over the Gaussian approach with state-of-the-art uncertainty modeling results in speech-based emotion recognition, along with an optimal and even faster convergence.

\end{abstract}


\section{Introduction}

Emotions can be inner subjective experiences, but in order to become socially relevant, they need to be expressed in social context (e.g., \cite{van2009emotions}). Therefore, emotions are typically studied as emotional expressions that others subjectively perceive and respond to \cite{Schuller2018-xi}. A common theoretical backdrop for analyzing emotions is the two-dimensional pleasure and arousal framework \cite{russell1980circumplex}, which describes emotional expressions in two continuous, bipolar, and orthogonal dimensions: pleasure-displeasure (\emph{valence}) and activation-deactivation (\emph{arousal}). One way in which emotions become expressed in social interactions, and therefore accessible for social signal processing (SSP), concerns speech signals. Speech emotion recognition (SER) research spans roughly two decades \cite{Schuller2018-xi}, with ever improving state-of-the-art results. As a consequence, affective sciences and SER has shown increasing prominence in high-critical and socially relevant domains, e.g. health, security, and employee well-being \cite{Schuller2018-xi, dukes2021rise, sridhar2021generative}.

A crucial challenge when studying emotional expressions and trying to establish a ground truth using the pleasure-arousal framework concerns the significant degree of subjectivity surrounding the perceptions of these expressions \cite{Schuller2018-xi}. Commonly, majority voting \cite{busso2008iemocap} or evaluator-weighted mean (EWE) \cite{grimm2005evaluation} have been used as approximations to obtain ground-truth labels. However, in the context of reliable real-world applications, it is required for SER systems to not only model ground-truth labels but also account for subjectivity based label uncertainty \cite{Schuller2018-xi, gunes2013categorical}. 

\begin{figure}[t!]
    \centering
    \includegraphics[width=0.48\textwidth]{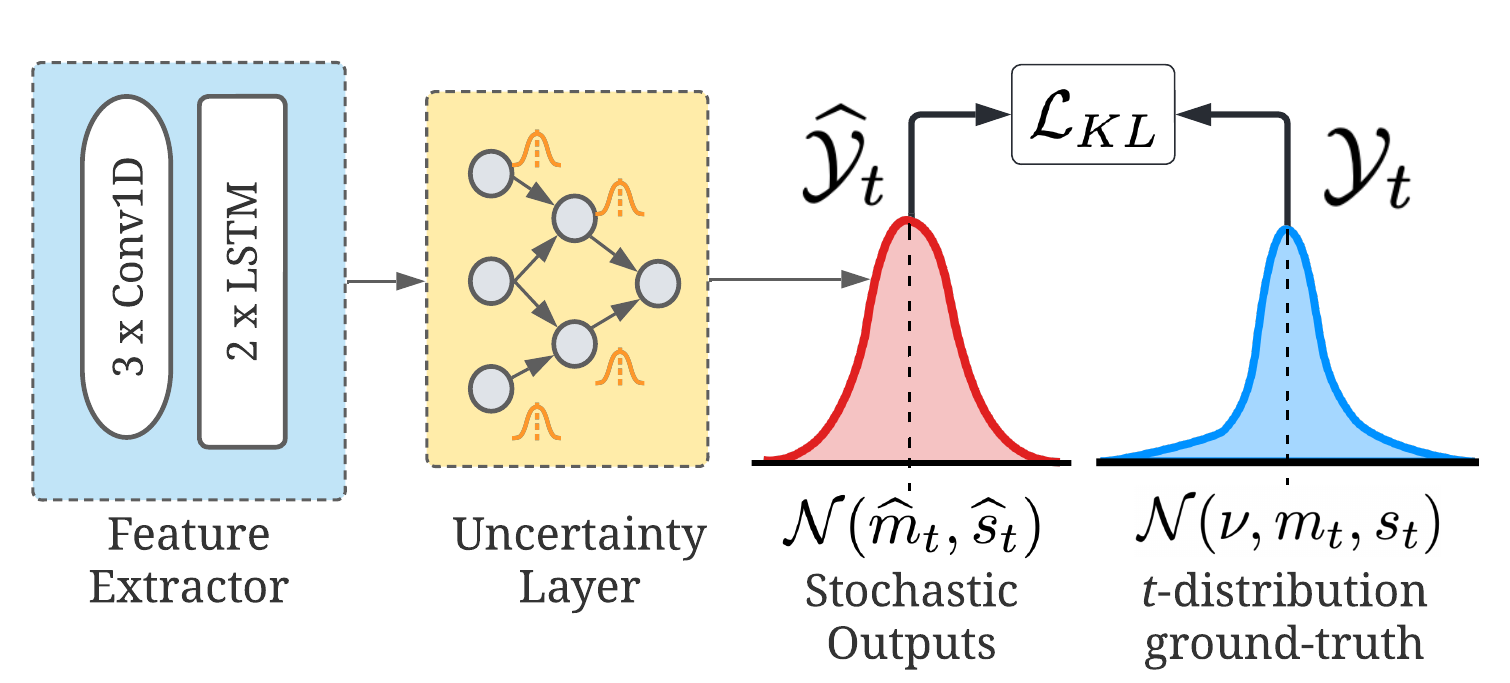}
    \caption{Overview of the proposed architecture and loss $\mathcal{L}_{\text{KL}}$.}
    \label{Fig:speechEmoBnn}
\end{figure}

In SER, label uncertainty has been approached using soft-labels \cite{sridhar2021generative}, multi-task learning (MTL) \cite{han2017hard, han2020exploring}, stochastic models \cite{sridhar2020modeling, t21_interspeech}, and label distribution learning \cite{foteinopoulou2021estimating, Prabhu2021EndToEndLU}. The subjective annotations of emotion creates a label distribution which explains the uncertainty in emotions \cite{sridhar2021generative}. In this light, label distribution learning techniques for label uncertainty in SER are gaining research focus, with improved performances \cite{foteinopoulou2021estimating, Prabhu2021EndToEndLU}. However, a problem with label distribution learning in SER is the limited annotations available \cite{Prabhu2021EndToEndLU, sridhar2021generative}, due to resource inefficient task of gaining more annotations \cite{sridhar2021generative}.




Emotion annotations as label distributions are usually modeled by making a Gaussian assumption on them for mathematical convenience \cite{foteinopoulou2021estimating, Prabhu2021EndToEndLU}. However, a Gaussian assumption with limited samples is not well justified \cite{kotz2004multivariate, bishop2006pattern}, as the central limit theorem (CLT) which primarily backs Gaussian distributions does not hold with insufficient numbers of samples \cite{ctl}. Publicly available SER datasets commonly comprise of only three to six annotations \cite{MspConv, MspPod, recolaDB, kossaifi2019sewa, raj2020defining}, and well agree that gaining more annotations is resource inefficient \cite{gatica-perez_automatic_2009, sridhar2021generative}. In this light, it is important for machine learning (ML) models to account for the limited annotations and model the label distribution accordingly. Alternately, Student's $t$-distribution, also known as {$t$-distribution}, is a probability distribution that also accounts for the number of samples available while modeling \cite{kotz2004multivariate}. Noting this, and the resource constraints in gaining more annotations, the $t$-distribution becomes a more appropriate choice for modeling emotion annotations.

In machine learning, two types of uncertainty can be distinguished. \textit{Label uncertainty} captures data inherent noise whereas \textit{model uncertainty} accounts for the uncertainty in model parameters \cite{zheng2021uncertainty}. Stochastic and probabilistic models have mainly been deployed for uncertainty modeling \cite{kohl2018probabilistic, garnelo2018conditional, blundell2015weight}. Bayes by Backpropagation (BBB) for Bayesian neural networks (BNN) \cite{blundell2015weight} uses \emph{simple gradient updates} to optimize weight distributions for \emph{stochastic outputs}, thereby are promising candidates for label distribution learning in SER.  



In this paper, we propose to model emotion annotations as a $t$-distribution, in contrast to a Gaussian assumption \cite{foteinopoulou2021estimating, Prabhu2021EndToEndLU}. To the best of our knowledge, this is the first time the problem of limited emotion annotations is tackled from an ML perspective, a common challenge in affective computing and SSP \cite{gatica-perez_automatic_2009}. For this, we adopt a BBB-based stochastic uncertainty model, as proposed in \cite{Prabhu2021EndToEndLU}, to include a $t$-distribution instead of a Gaussian. To this end, we introduce a Kullback-Leibler  (KL) divergence loss for label uncertainty that quantifies distribution similarity between stochastic emotion predictions, modeled as a Gaussian distribution, and \emph{ground-truth emotion annotations}, modeled as a \emph{$t$-distribution}. Subsequently, we present analyses to reveal the benefits of using $t$-distribution over a Gaussian. Finally, we show that the BBB-based uncertainty model trained on the proposed $t$-distribution based KL-divergence loss can aptly capture label uncertainty with state-of-the-art results, along with a robust loss curve.



\section{Related Work}
\subsection{Ground-truth labels}
To handle subjectivity in emotional expressions, annotations $\{y_{1}, y_{2}, .., y_{a}\}$ for emotions are collected from $a$ annotators \cite{recolaDB, raj2020defining}. The \emph{ground-truth label} is then obtained as the mean $m$ over all annotations from $a$ annotators \cite{abdelwahab2019active, avec16},
\begin{equation} \label{eq:mean-annot}
    {m} = \dfrac{1}{a} \sum_{i=1}^a y_{i}.
\end{equation}
Alternatively, the EWE, which weights annotations with inter-annotator correlations, has been proposed and referred to as the \emph{gold-standard} $\widetilde{m}$ \cite{grimm2005evaluation}. Both $m$ and $\widetilde{m}$ based approximation of ground-truth leads to loss of information on subjectivity \cite{sridhar2021generative}.




Traditional SER approaches, given a raw audio sequence of $T$ frames $\mathcal{X} = [x_1, x_2, ..., x_T]$, aim to estimate either the $m_t$ or $\widetilde{m}_t$ for each time frame $t \in [1,T]$, referred to as $\widehat{m}_t$. The concordance correlation coefficient (CCC) has been widely used as a loss function for this task \cite{Schuller2018-xi}. For Pearson correlation $r$, the CCC between $m$ and $\widehat{m}$, for $T$ frames, is formulated as
\begin{equation}\label{loss:CCC}
    \mathcal{L}_{\text{CCC}}(m) = {\frac {2r \sigma_{m}\sigma_{\widehat{m}}}{\sigma _{m}^{2}+\sigma_{\widehat{m}}^{2}+(\mu _{m}-\mu _{\widehat{m}})^{2}}},
\end{equation}
where $\mu_{m} = \dfrac{1}{T} \sum_{t=1}^{T}m_t$, $\sigma_{m}^{2} = \dfrac{1}{T} \sum_{t=1}^T (m_t - \mu_m)^2$, and $\mu _{\widehat{m}}$, $\sigma_{\widehat{m}}^{2}$ are obtained similarly for $\widehat{m}$. 

\subsection{Label uncertainty in SER}
Alternative to exclusively modeling $m_t$ or $\widetilde{m}_t$, works have attempted to model ground truth that also explains inter-annotator disagreement, for example by means of  soft labels \cite{sridhar2021generative} and entropy of disagreement \cite{steidl2005all}. Fayek et al. \cite{fayek2016modeling} and Tarantino et al. \cite{Tarantino2019SelfAttentionFS} proposed to learn soft labels instead of ${m}_t$ with improved performance. Steidl et al. \cite{steidl2005all} quantified label uncertainty using the entropy measure, and trained a model to minimize the difference in entropy between model outputs and annotator disagreement. Sridhar et al. \cite{sridhar2021generative} proposed an auto-encoder based learning technique to jointly model soft- and hard-labels of emotion annotations, and subsequently estimating label uncertainty as the entropy on soft-labels. 

Label uncertainty has also been approached as a prediction task by estimating either the moments of the distribution \cite{han2017hard, han2020exploring} or the distribution in itself \cite{foteinopoulou2021estimating, Prabhu2021EndToEndLU}. Han et al. \cite{han2017hard, han2020exploring} used an MTL approach to model the unbiased standard deviation $s$ of $a$ annotators as an auxiliary task,
\begin{equation}\label{eq:pu}
    s = \sqrt{\dfrac{1}{a - 1} \sum_{i=1}^a (y_i - m)^2}.
\end{equation}
Sridhar et al. \cite{sridhar2020modeling} introduced a Monte-Carlo dropout model to obtain uncertainty estimates from the distribution of stochastic outputs. Foteinopoulou et al. \cite{foteinopoulou2021estimating} trained a MTL network using a KL divergence loss that models emotion annotations as a uni-variate Gaussian with mean $m$ and unknown variance. Raj Prabhu et al. \cite{Prabhu2021EndToEndLU} introduced a stochastic BNN and trained them on Gaussian emotion annotations. Notwithstanding their improved performances, in \cite{foteinopoulou2021estimating, Prabhu2021EndToEndLU} \emph{Gaussian} emotion annotations are assumed, despite only having limited annotations. Apart from the apparent mathematical incorrectness of this assumption, they are susceptible to unreliable $m$ and $s$ for lower values of $a$ and sparsely distributed annotations \cite{Prabhu2021EndToEndLU}.




\subsection{On distributions}

A Gaussian distribution $\widehat{\mathcal{Y}}\sim {\mathcal {N}}(\widehat{\mu} ,\widehat{\sigma}^{2})$ is a continuous probability distribution for a real-valued random variable $y$, with general form of its probability density function \cite{bishop2006pattern}
\begin{equation}\label{pdf:gaussian}
p(y\mid \widehat{\mu}, \widehat{\sigma})={\frac {1}{\widehat{\sigma} {\sqrt {2\pi }}}}e^{-{\frac {1}{2}}\left({\frac {y-\widehat{\mu}}{\widehat{\sigma} }}\right)^{2}}.
\end{equation}
The parameters $\widehat{\mu}$ and $\widehat{\sigma}$ are the mean and standard deviation of the distribution, respectively. 

Due to its simplicity and intelligibility, Gaussian distributions are often used to model random variables whose distribution are unknown \cite{blundell2015weight, foteinopoulou2021estimating, Prabhu2021EndToEndLU}. Their importance is however backed by the CLT which only holds \textit{as the number of observations of the random variable grows} \cite{ctl}.
However, due to the resource constraints in collecting annotations, in most human-behaviour research \cite{gatica-perez_automatic_2009} and in SER \cite{recolaDB, MspConv, MspPod, kossaifi2019sewa}, we do not have sufficient annotations to assume a Gaussian distribution on them. As this is a common challenge for reliable real-world applications, it is important for SER algorithms to account for \emph{limited} annotations and model annotation distributions accordingly. Kotz and Nadarajah \cite{kotz2004multivariate}, and, Bishop and Nasrabadi \cite{bishop2006pattern}, note that in scenarios of limited observations and samples the $t$-distribution becomes more robust and realistic over a Gaussian.






Student's $t$-distribution is a probability distribution that arises when estimating the moments of a normally distributed population in \emph{situations where the sample size is small} \cite{kotz2004multivariate, walpole2017probability}, with the probability density function given by \cite{kruschke2015doing, villa2014objective, villa2018objective},
\begin{equation}\label{pdf:tstudent}
    p(y\mid \nu ,{{\mu }},{{\sigma }})=\frac{1}{\mathrm{B} (\frac{1}{2}, \frac{\nu}{2})} \frac{1} {\sqrt{\nu{\sigma }^2}} \left(1+\frac{(y-{\mu })^2}{\nu{\sigma }^2}\right)^{-{\frac {\nu +1}{2}}},
\end{equation}
where $\nu$ denotes the degrees of freedom and B(., .) is the Beta function, for Gamma function $\Gamma$, formulated as,
\begin{equation}
    B(i, j) = \dfrac{\Gamma(i)\,\Gamma(j)}{\Gamma(i+j)}.
\end{equation}
The density function \eqref{pdf:tstudent} resembles the bell shape of a normally distributed variable, except that it has heavier tails, meaning that it better captures values that fall far from its mean \cite{bishop2006pattern, kotz2004multivariate}. The degree of freedom $\nu$, also known as the normality parameter, controls the normality of the distribution, and is correlated with the ${\sigma}$ parameter \cite{kotz2004multivariate, bishop2006pattern}. The standard deviation ${\sigma}$ in \eqref{pdf:tstudent} is scaled by $\nu$ and is formulated as 
\begin{equation}\label{eq:scale-sigma}
    {\sigma } = {{\sigma } \, \sqrt{\frac {\nu }{\nu -2}} {\text{  for }}}\nu >2.  
\end{equation}
As $\nu$ increases, the $t$-distribution approaches the normal distribution \cite{villa2018objective}. 


\section{Proposed $t$-distribution Label Uncertainty Model}

To better represent subjectivity in annotations of emotional expressions, we propose to estimate the \emph{emotion annotation distribution} $\mathcal{Y}_t$ for each frame $t$. For this, in contrast to a Gaussian assumption $\mathcal{Y}_t \sim \mathcal{N}(m_t, s_t)$ \cite{foteinopoulou2021estimating, Prabhu2021EndToEndLU}, we model the annotations as a $t$-distribution with degrees of freedom $\nu$:
\begin{equation}
\mathcal{Y}_t \sim \mathcal{N}(\nu, m_t, s_t).
\end{equation}
Thus, the goal is to obtain an estimate $\widehat{\mathcal{Y}}_t$ of $\mathcal{Y}_t$. 





\subsection{Model architecture}


We adopt an end-to-end architecture, initially proposed by Raj Prabhu et al. \cite{Prabhu2021EndToEndLU}, which uses a feature extractor \cite{Tzirakis2018-speech} to learn temporal-paralinguistic features from $x_t$, and a BBB-based uncertainty layer \cite{blundell2015weight} to estimate $\mathcal{Y}_t$. We include the $t$-distribution modeling as part of the architecture, and the architecture proposed here can be seen in Figure \ref{Fig:speechEmoBnn}.




Unlike a standard neuron which optimizes a deterministic weight $w$, the BBB-based neuron learns a probability distribution on the weight $P(w|\mathcal{D})$, parametrized by $\theta=(\mu_w, \sigma_w)$ using a Gaussian $\mathcal{N}(\mu_w, \sigma_w)$, given the training data $\mathcal{D}$ \cite{blundell2015weight}. For an optimized $\theta$, the predictive distribution $\widehat{\mathcal{Y}}_t$ for an audio frame $x_t$, is given by $P (\widehat{y}_t|x_t) = \mathbb{E}_{P(w|D)}[P(\widehat{y}_t|x_t, w)]$, where $\widehat{y}_t$ are realizations of $\widehat{\mathcal{Y}}_t$. Stochastic outputs in BBB are achieved using multiple forward passes $n$ with stochastically sampled weights $w$, thereby modeling $\widehat{\mathcal{Y}}_t$ using the $n$ stochastic estimates. Following \cite{blundell2015weight}, the BBB-based MLP is trained on the negative evidence lower bound (ELBO),
\begin{equation}\label{loss:BBB-stoch}
    \mathcal{L}_{\text{BBB}} \approx \sum_{i=1}^{\textit{$n$}} \log q(w^{(i)}|\theta) - \log P(w^{(i)}) - \log P(D|w^{(i)}),
\end{equation}
where $q(w|\theta)$ is the variational posterior that minimizes the KL divergence with the true Bayesian posterior, and $w^{(i)}$ is the $i^{th}$ sampled weight from $q(w|\theta)$. Finally, as suggested in \cite{Prabhu2021EndToEndLU}, during testing the uncertainty estimate $\widehat{s}_t$ is the standard deviation of $\widehat{\mathcal{Y}}_t$, and mean estimate $\widehat{m}_t$ is the realization $\widehat{y}_t$ obtained using the mean of the optimized weights $\mu_w$.



\subsection{$t$-distribution label uncertainty loss derivation}
To capture the label uncertainty, we derive a KL divergence based loss function, between the Gaussian stochastic outputs $\widehat{\mathcal{Y}}$ and the $t$-distribution ground-truth $\mathcal{Y}$. Note that assuming a Gaussian distribution on the stochastic outputs $\widehat{\mathcal{Y}}$ is a fair assumption as the number of stochastic outputs to model $\widehat{\mathcal{Y}}$ can be controlled using $n$ in \eqref{loss:BBB-stoch}. As the number of sample observations for a distribution approaches thirty a $t$-distribution converges to a stable Gaussian \cite{villa2014objective, villa2018objective}. Noting this, we intend to choose a $n$ greater than 30 and thereby assume $\widehat{\mathcal{Y}}$ to be Gaussian. As a positive side effect, we result in deriving the KL divergence between a Gaussian and a $t$-distribution, in contrast to between two $t$-distributions, with the later involving mathematical complexities in calculating intractable expectations for a loss function.

\begin{figure*}[t!]
     \captionsetup[subfigure]{justification=centering}
     \centering
     \begin{subfigure}[b]{0.245\textwidth}
            \includegraphics[width=\textwidth]{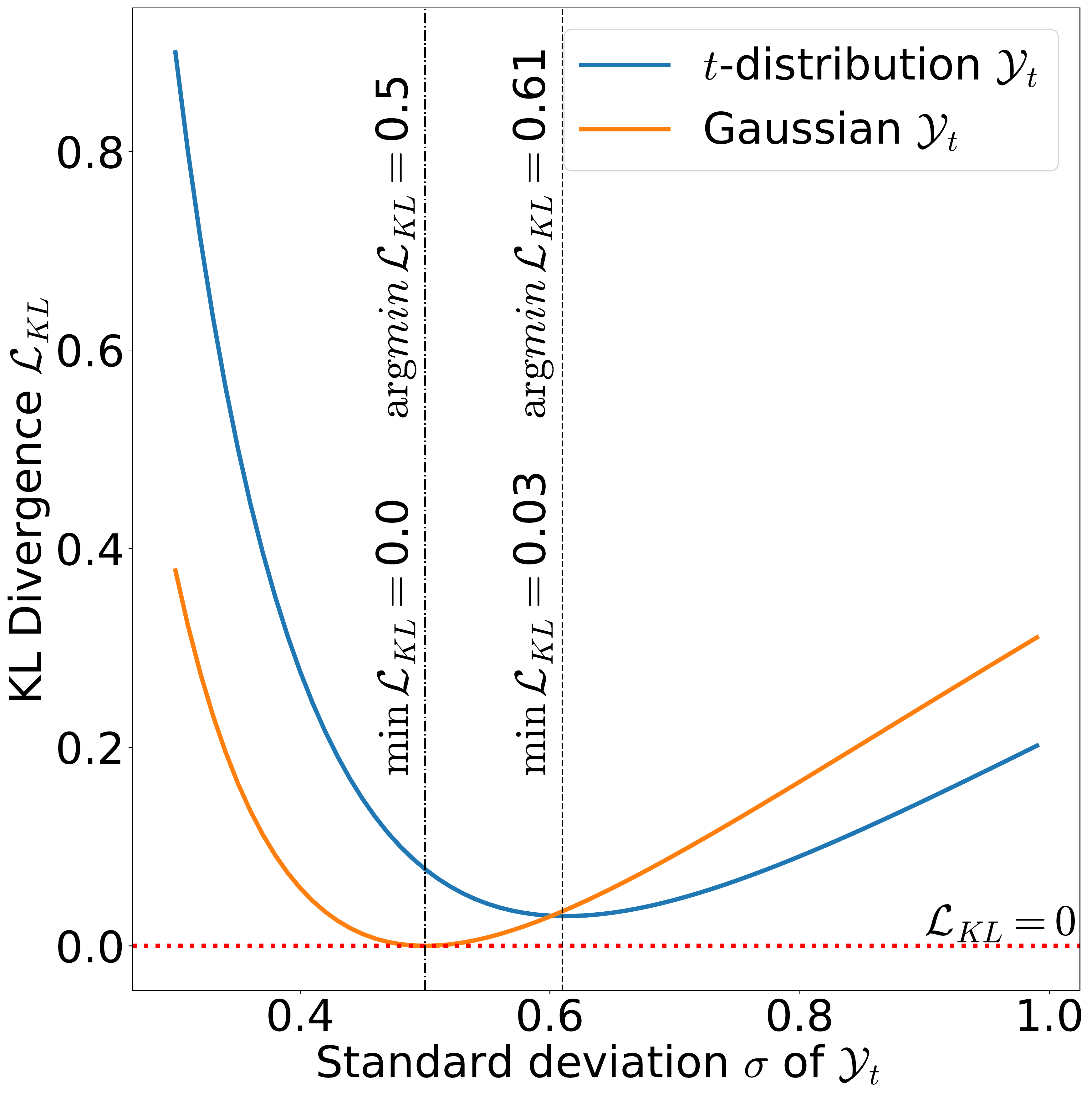}%
         \caption{$\widehat{\mathcal{Y}}_t \sim \mathcal{N}(0, 0.5)$, and $\nu=6$}
      \label{fig:kl-analysis-sigma5_nu6}%
     \end{subfigure}
     \hfill
     \begin{subfigure}[b]{0.245\textwidth}
            \includegraphics[width=\textwidth]{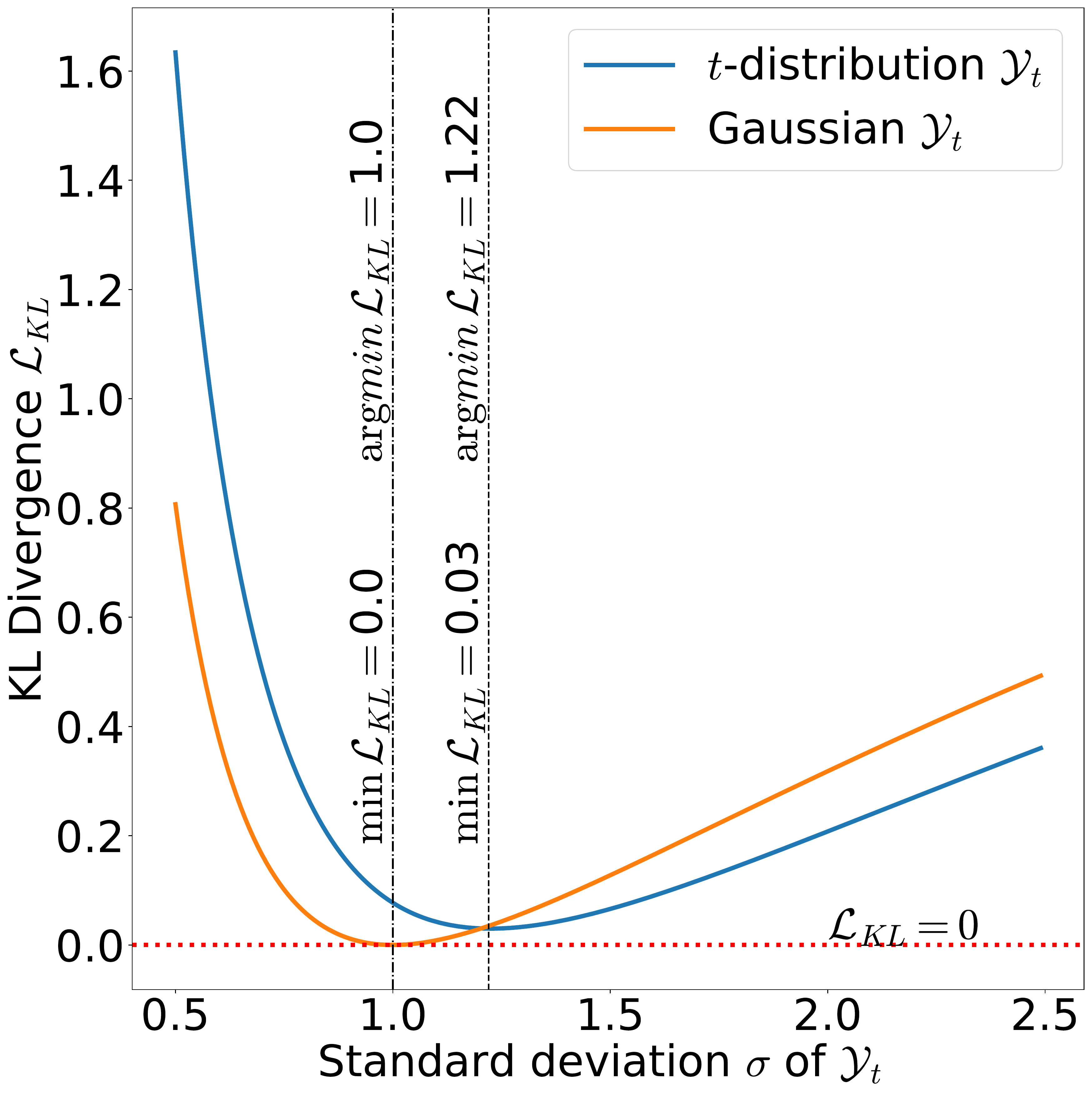}%
         \caption{$\widehat{\mathcal{Y}}_t \sim \mathcal{N}(0, 1)$, and $\nu=6$}
      \label{fig:kl-analysis-sigma1_nu6}%
     \end{subfigure}
     \hfill
     \begin{subfigure}[b]{0.245\textwidth}
            \includegraphics[width=\textwidth]{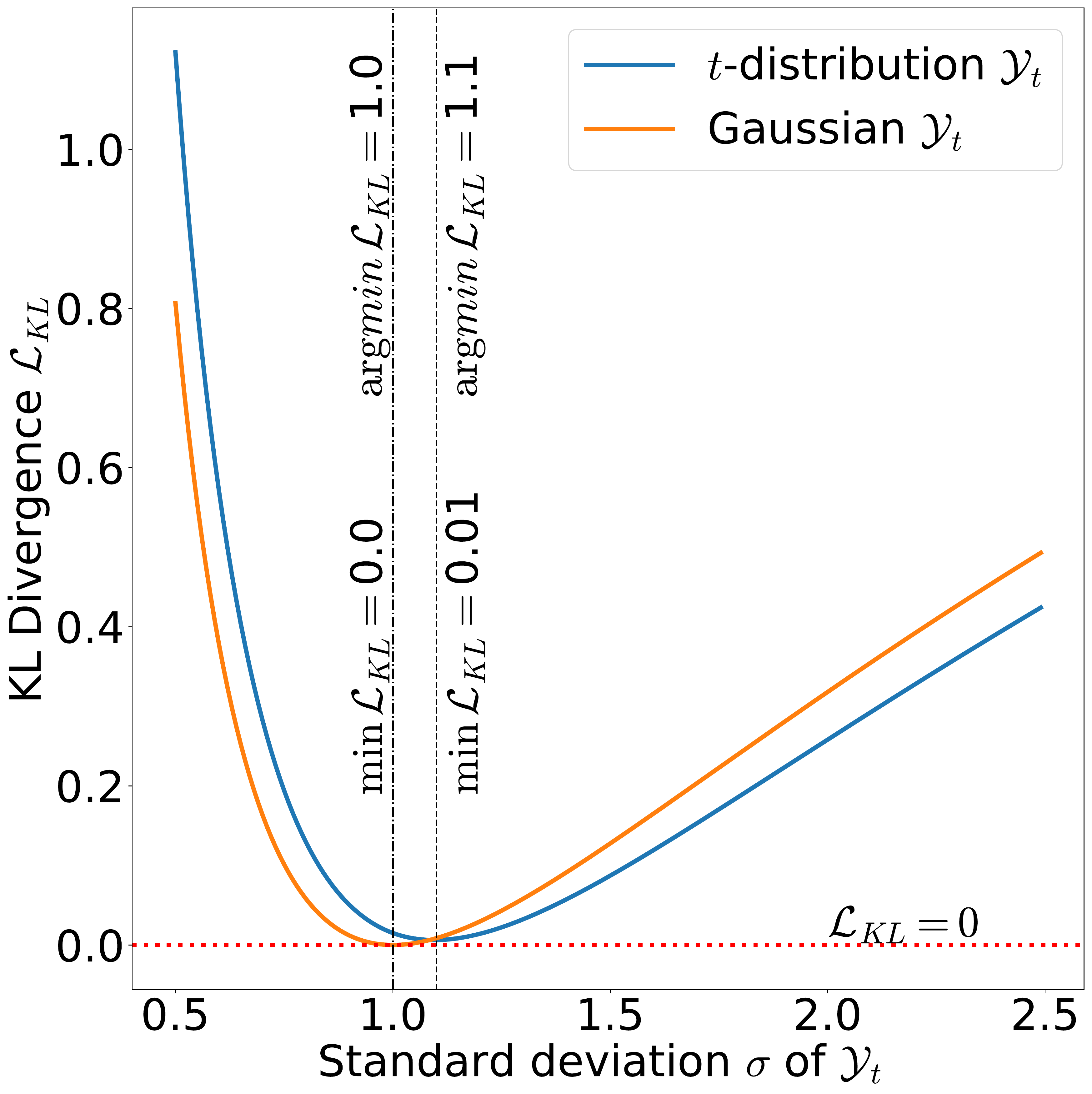}%
         \caption{$\widehat{\mathcal{Y}}_t \sim \mathcal{N}(0, 1)$, and $\nu=12$}
      \label{fig:kl-analysis-sigma1_nu12}%
     \end{subfigure}
     \hfill
     \begin{subfigure}[b]{0.245\textwidth}
            \includegraphics[width=\textwidth]{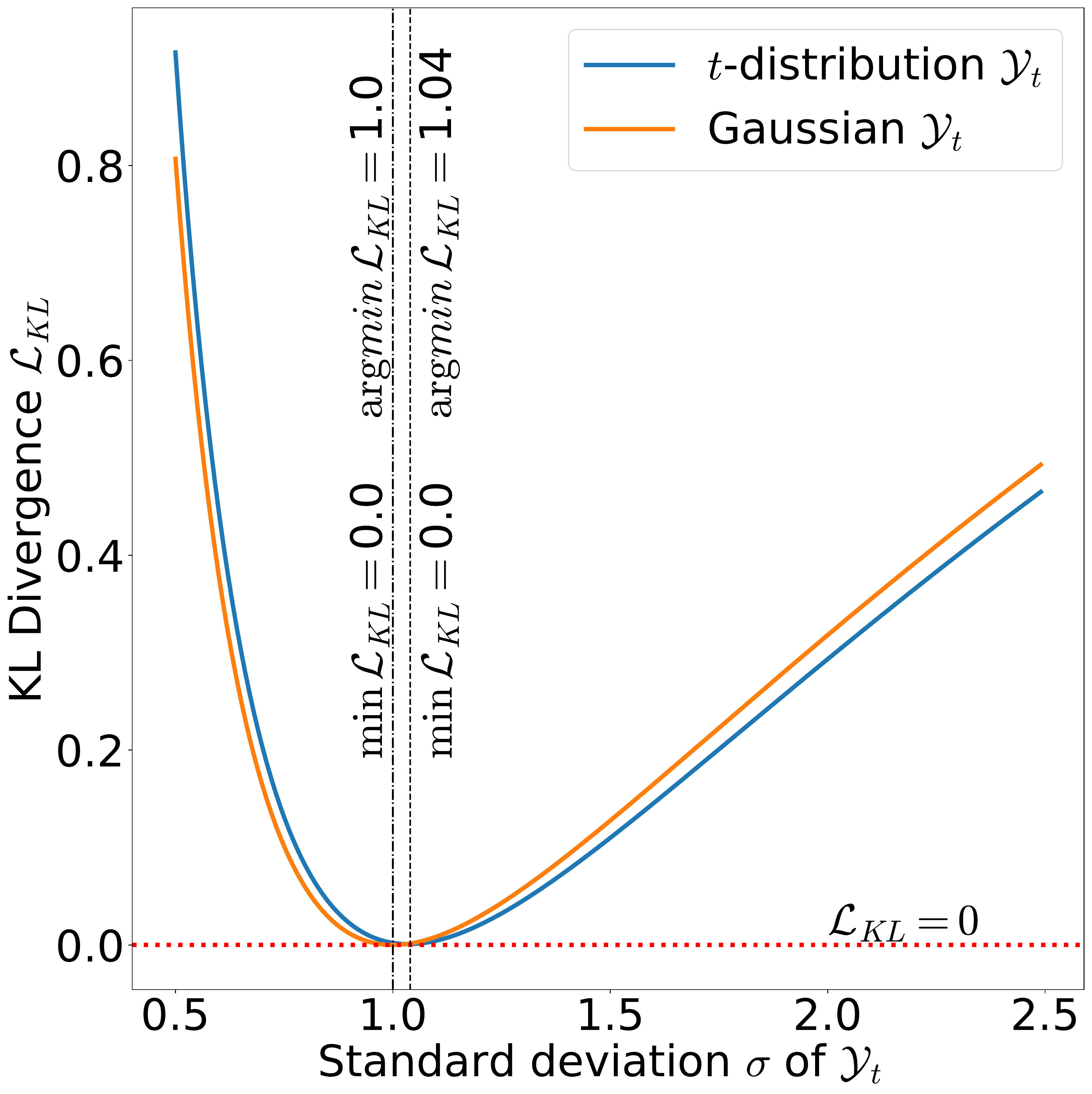}%
         \caption{$\widehat{\mathcal{Y}}_t \sim \mathcal{N}(0, 1)$, and $\nu=30$}
         \label{fig:kl-analysis-sigma1_nu30}
     \end{subfigure}
        \caption{Analysis of the $t$-distribution based KL divergence $\mathcal{L}_{KL}$ \eqref{loss:final-KL}, in comparison with Gaussian $\mathcal{L}_{KL}$  \eqref{eq:Gauss-KL}.}
        \label{fig:kl-analysis}
\end{figure*}

For a Gaussian $\widehat{\mathcal{Y}}$ (see \eqref{pdf:gaussian}), and a $t$-distributed $\mathcal{Y}$ (see \eqref{pdf:tstudent}), the $\mathcal{L}_{KL}$ is formulated as \cite{kullback1951information, murphy2012machine},
\begin{equation}\label{loss:KL}
    \mathcal{L}_{KL} = f_{KL}({\mathcal{Y}}_t\| \widehat{\mathcal{Y}}_t) = H(\mathcal{Y}_t, \widehat{\mathcal{Y}}_t)- H({\mathcal{Y}}_t),
\end{equation}
where $\text{H}(.,.)$ is the cross-entropy between two distributions, and $\text{H}(.)$ is the entropy of a distribution. Similar to \cite{Prabhu2021EndToEndLU}, in \eqref{loss:KL}, we choose the true distribution $\mathcal{Y}_t$ to precede its estimate $\widehat{\mathcal{Y}}_t$, promoting a mean-seeking approximation rather than a mode-seeking one and capturing the full distribution \cite{Goodfellow-et-al-2016}.

 
The cross-entropy term $\text{H}(. , .)$ in \eqref{loss:KL}, using \eqref{pdf:gaussian}, can be further formulated as,
\begin{eqnarray}
    &    \lefteqn{\text{H} ({\mathcal{Y}}_t, \widehat{\mathcal{Y}}_t) = - \int {\mathcal{Y}}_t(y)\,\, \log\,\widehat{\mathcal{Y}}_t(y)\, dy } \notag
        \\ 
    &    =& - \int \mathcal{Y}_t(y)\,\, \Big[ \log\Big( \frac{1}{\sqrt{2\pi\widehat{\sigma}^2}}e^{-{\frac {1}{2}}\left({\frac {y-\widehat{\mu}}{\widehat{\sigma}}}\right)^{2}} \Big) \Big] dy   \notag
        \\
    &    =& - \int \mathcal{Y}_t(y)\,\, \Big[ -\frac{1}{2} \log(2\pi\widehat{\sigma}^2) + \log\Big( e^{-{\frac {1}{2}}\left({\frac {y-\widehat{\mu} }{\widehat{\sigma}}}\right)^{2}} \Big)\Big] dy  \notag
        \\
    &    =& \dfrac{1}{2} \log(2\pi\widehat{\sigma}^2) + \int \mathcal{Y}_t(y)\,\, \Big( \frac{(y-\widehat{\mu})^2}{2\widehat{\sigma}^2} \Big) dy  \notag
        \\
    &    =& \dfrac{1}{2} \log(2\pi\widehat{\sigma}^2) \,  +  \,  \frac{1}{2\widehat{\sigma}^2} \Big[ \int \mathcal{Y}_t(y) y^2\, dy - 2\widehat{\mu} \int \mathcal{Y}_t(y)\, y\, dy \notag
    \\
    && \qquad\qquad\qquad\qquad\qquad\qquad  + \, \widehat{\mu}^2 \int \mathcal{Y}_t(y)\, dy \Big] \label{eq:init-cross-entropy}
\end{eqnarray}

Noting that $\int \mathcal{Y}_t(y) y^2\, dy = {\mu}^2 + {\sigma}^2$, $\int \mathcal{Y}_t(y)\, y\, dy = {\mu}$, and $\int \mathcal{Y}_t(y)\, dy = 1$, where ${\mu}$ and ${\sigma}$ are parameters of the $t$-distribution $\mathcal{Y}_t$, $p(y\mid \nu, {\mu}, {\sigma})$, the equation \eqref{eq:init-cross-entropy} becomes,

\begin{eqnarray}
    &\lefteqn{} =& \dfrac{1}{2} \log(2\pi\widehat{\sigma}^2) \,\, + \,\, \frac{1}{2\widehat{\sigma}^2} \Big[ {\sigma}^2 \,+\, {\mu}^2 \,-\, 2\widehat{\mu}\mu \,+\, \widehat{\mu}^2 \Big] \notag
    \\
    & \lefteqn{} =& \dfrac{1}{2} \log(2\pi\widehat{\sigma}^2) \,\, + \,\, \frac{{\sigma}^2 \,+\, ({\mu}-\widehat{\mu})^2}{2\widehat{\sigma}^2} \label{eq:final-crossentropy}
\end{eqnarray}
        
        

Finally, using \eqref{eq:final-crossentropy} in \eqref{loss:KL}, our proposed KL divergence is
\begin{equation}\label{loss:final-KL}
\boxed{
    \mathcal{L}_{KL} = \frac{1}{2} \log(2\pi\widehat{\sigma}^2) \,\, + \,\, \frac{{\sigma}^2 \,+\, ({\mu}-\widehat{\mu})^2}{2\widehat{\sigma}^2}  - H({\mathcal{Y}}_t) .
    }
\end{equation}
We implement \eqref{loss:final-KL} as a custom loss function using the pytorch package \cite{paszke2019pytorch}, by extending the \texttt{studentT} sub-package\footnote{Code for the models and the loss functions introduced are available at \,\,\,\,\,\url{https://github.com/sp-uhh/label-uncertainty-ser}}. 

 



Similarly, as used in \cite{Prabhu2021EndToEndLU}, the KL divergence between two Gaussians $\mathcal{N}({\mu},\sigma^{2})$ and $\mathcal{N}(\widehat{\mu},\widehat{\sigma}^{2})$ is given by \cite{bishop2006pattern, paszke2019pytorch}
\begin{equation}\label{eq:Gauss-KL}
    \mathcal{L}_{KL} = \log\left(\frac{\widehat{\sigma}}{{\sigma}}\right) + \,\, \frac{{\sigma}^2 \,+\, ({\mu}-\widehat{\mu})^2}{2\widehat{\sigma}^2} - \frac {1}{2}.
\end{equation}

While the two loss-functions \eqref{eq:Gauss-KL} and \eqref{loss:final-KL} have their second term in common, two differences can be noted. Firstly, as \eqref{eq:Gauss-KL} calculates the divergence between two similar distributions, $\mathcal{Y}_t$ and $\widehat{\mathcal{Y}}_t$, \eqref{eq:Gauss-KL} includes the logarithm of the ratio between the two Gaussian's standard deviation in its formulation. However, in \eqref{loss:final-KL}, the deviations of $\widehat{\mathcal{Y}}_t$ and ${\mathcal{Y}}_t$ are separately quantified using terms $\frac{1}{2} \log(2\pi\widehat{\sigma}^2)$ and $H({\mathcal{Y}}_t)$, respectively. Secondly, the number of annotators is included in \eqref{loss:final-KL} by scaling ${\sigma}_t$ with normality factor $\nu$, using \eqref{eq:scale-sigma}. The implication of these differences, and the quantitative differences between \eqref{eq:Gauss-KL} and \eqref{loss:final-KL} are presented and analyzed in the following section.

\subsection{$t$-distribution loss analysis} \label{Section:kl-analysis}

In contrast to \cite{Prabhu2021EndToEndLU}, where the loss function is $\mathcal{L}_{KL}$ between two Gaussians, and to \cite{foteinopoulou2021estimating}, where $\mathcal{L}_{KL}$ is between a Gaussian and a Dirac delta, our proposed loss \eqref{loss:final-KL} formulates the KL divergence between a Gaussian and a $t$-distribution to capture label uncertainty when only limited annotations are available. 
To validate the derivation and to further understand the advantages of the $t$-distribution $\mathcal{L}_{KL}$ \eqref{loss:final-KL} over the Gaussian $\mathcal{L}_{KL}$ \eqref{eq:Gauss-KL}, we plot the $\mathcal{L}_{KL}$ values as a function of varying $\sigma$ of $\mathcal{Y}_t$, for \eqref{loss:final-KL} and \eqref{eq:Gauss-KL}. We perform this analysis under four different scenarios, by varying parameters $\widehat{\sigma}$ and ${\nu}$, i) Figure \ref{fig:kl-analysis-sigma5_nu6} for scenario $\widehat{\sigma}=0.5$ and $\nu=6$, ii) Figure \ref{fig:kl-analysis-sigma1_nu6} for scenario $\widehat{\sigma}=1.0$ and $\nu=6$, iii) Figure \ref{fig:kl-analysis-sigma1_nu12} for scenario $\widehat{\sigma}=1.0$ and $\nu=12$, and, iv) Figure \ref{fig:kl-analysis-sigma1_nu30} for scenario $\widehat{\sigma}=1.0$ and $\nu=30$.


From Figure \ref{fig:kl-analysis}, firstly, we see that $\mathcal{L}_{KL}$ behaves differently when the ground-truth $\mathcal{Y}_t$ is modeled as a $t$-distribution \eqref{loss:final-KL}, in comparison to the Gaussian assumption \eqref{eq:Gauss-KL}. Specifically, from Figure \ref{fig:kl-analysis-sigma5_nu6}, for $\widehat{\sigma}=0.5$ and $\nu=6$, we see that the minimum $\mathcal{L}_{KL}$ \eqref{loss:final-KL} is achieved only at ${\sigma}=0.61$, in contrast to the Gaussian \eqref{eq:Gauss-KL} $\widehat{\sigma}={\sigma}=0.5$. While the Gaussian attempts exactly fitting the model to the ground-truth ${\sigma}=0.5$, the $t$-distribution tries to fit on a more relaxed ${\sigma}=0.61$ by also considering the reduced degree of freedom $\nu=6$. This behaviour is similar to that observed during the confidence intervals calculation using a Gaussian and $t$-distribution \cite{rees2001book}, where a $t$-distribution shows relaxation on $\sigma$ with respect to $\nu$. Moreover, Bishop and Nasrabadi \cite{bishop2006pattern} associate this relaxed ${\sigma}$ towards the increased robustness of the $t$-distribution to outliers and sparse distributions.


Secondly, we note that the observed relaxation on ${\sigma}$ is dependent on two factors, 1) the standard-deviation of the stochastic outputs $\widehat{\sigma}$, and 2) the degree of freedom of the ground-truth $\nu$. From figures \ref{fig:kl-analysis-sigma5_nu6} and \ref{fig:kl-analysis-sigma1_nu6}, we see that, while $\nu$ is constant, the relaxation on $\sigma$ \emph{increases} along with an increase in $\widehat{\sigma}$. At $\widehat{\sigma}=0.5$ a relaxation of $0.11$ is made by $t$-distribution \eqref{loss:final-KL} from $0.5$ to $0.61$, while a larger relaxation of $0.22$ is made for $\widehat{\sigma}=1.0$. Similarly, from figures \ref{fig:kl-analysis-sigma1_nu12} and \ref{fig:kl-analysis-sigma1_nu30}, we see that, while $\widehat{\sigma}$ is constant, as $\nu$ increases the relaxation on $\sigma$ \emph{decreases}. That is, the $t$-distribution \eqref{loss:final-KL} starts behaving similar to that of the Gaussian, inline with literature that states that as the degree of freedom $\nu$ of $t$-distribution increases, the distribution converges into a Gaussian \cite{villa2014objective, villa2018objective, kotz2004multivariate}. This is also inline with our initial motivation behind using the $t$-distribution, which we expected to account for the number of annotators $\nu$ while fitting on annotation distribution $\mathcal{Y}$.

From a machine learning and SER perspective, from Figure \ref{fig:kl-analysis}, we note several benefits that $t$-distribution loss term $\mathcal{L}_{KL}$ \eqref{loss:final-KL} brings forth in-terms of label uncertainty modeling. Firstly, training on a $t$-distribution based $\mathcal{L}_{KL}$ \eqref{loss:final-KL} leads to training on a relaxed $s_t$, and thereby can lead to better capturing of the whole ground-truth label distribution. Moreover, the resulting loss function is mathematically more solid than a Gaussian assumption as in \cite{foteinopoulou2021estimating, Prabhu2021EndToEndLU}, when less than \emph{thirty} annotations are available. Secondly, we note that the $t$-distribution $\mathcal{L}_{KL}$ \eqref{loss:final-KL} values are always higher for lower values of $\sigma$ and $\widehat{\sigma}$, in all cases. This, in comparison to the Gaussian $\mathcal{L}_{KL}$ \eqref{eq:Gauss-KL}, might lead to larger penalization of the model through the $\mathcal{L}_{KL}$ loss, and may thereby promote better and quicker convergence during training. Finally, the $t$-distribution $\mathcal{L}_{KL}$ \eqref{loss:final-KL} also adapts to different datasets with different number of annotators by considering the number of annotations available during training. 


\subsection{Training loss}
The proposed uncertainty training loss is formulated as,
\begin{equation}\label{eq:end-to-end_loss}
   \mathcal{L} = (1 - \mathcal{L}_{\text{CCC}}(m)) + \mathcal{L}_{\text{BBB}} + \mathcal{L}_{\text{KL}}.
\end{equation}
 Intuitively, $\mathcal{L}_{\text{CCC}}(m)$ \eqref{loss:CCC} optimizes for mean predictions $m$, $\mathcal{L}_{\text{BBB}}$ \eqref{loss:BBB-stoch} optimizes for BBB weight distributions, and $\mathcal{L}_{\text{KL}}$ \eqref{loss:final-KL} optimizes for the label distribution $\mathcal{Y}_t$ as a $t$-distribution. 
 



\section{Experimental Setup}
\subsection{Dataset}
\begin{figure}[t!]
     \centering
     \begin{subfigure}[b]{0.24\textwidth}
        \includegraphics[width=\textwidth]{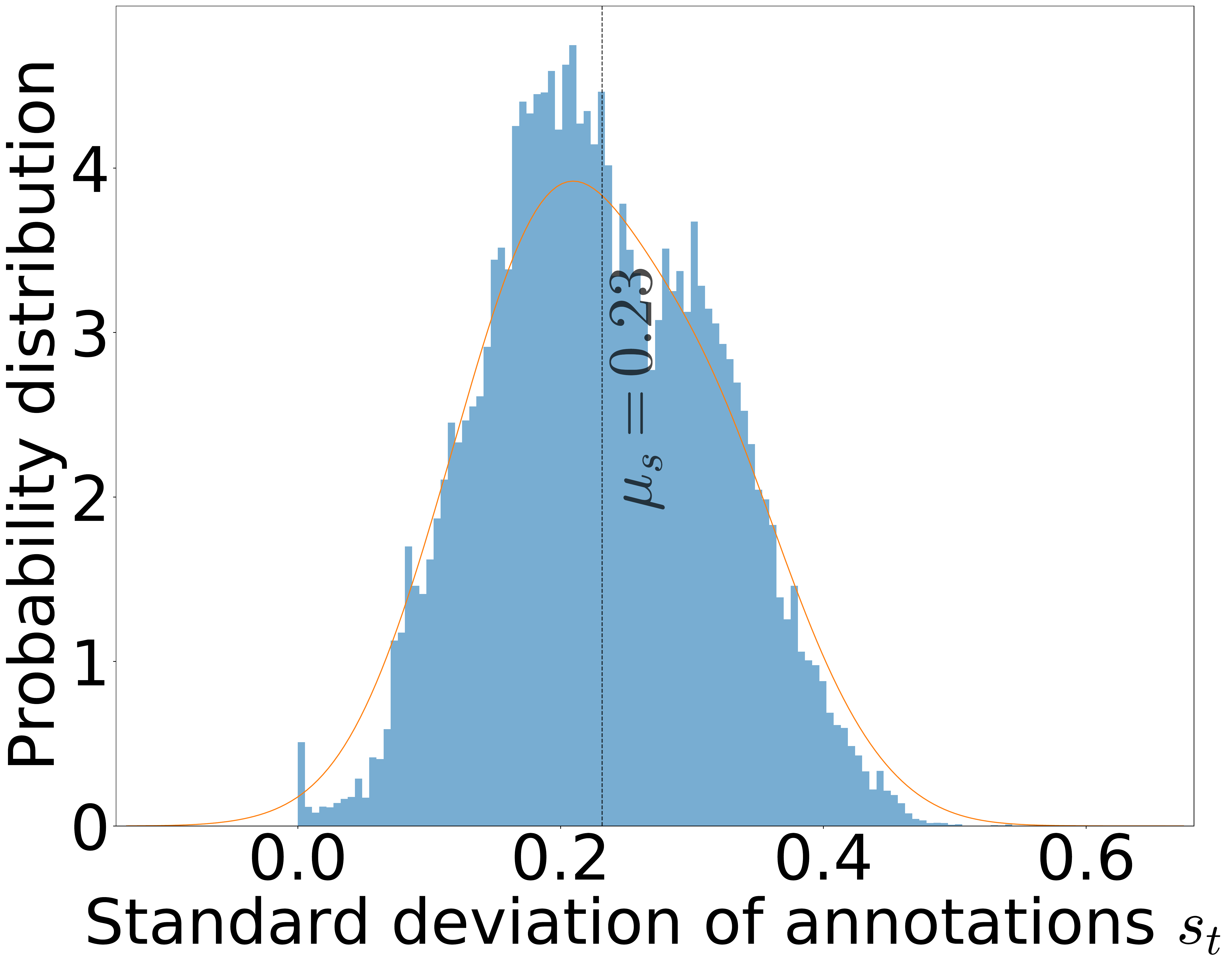}
        \caption{For arousal.}
        \captionsetup{justification=centering}
        \label{Fig:s_dist_arousal}
     \end{subfigure}
     \hfill
     \begin{subfigure}[b]{0.24\textwidth}
        \includegraphics[width=\textwidth]{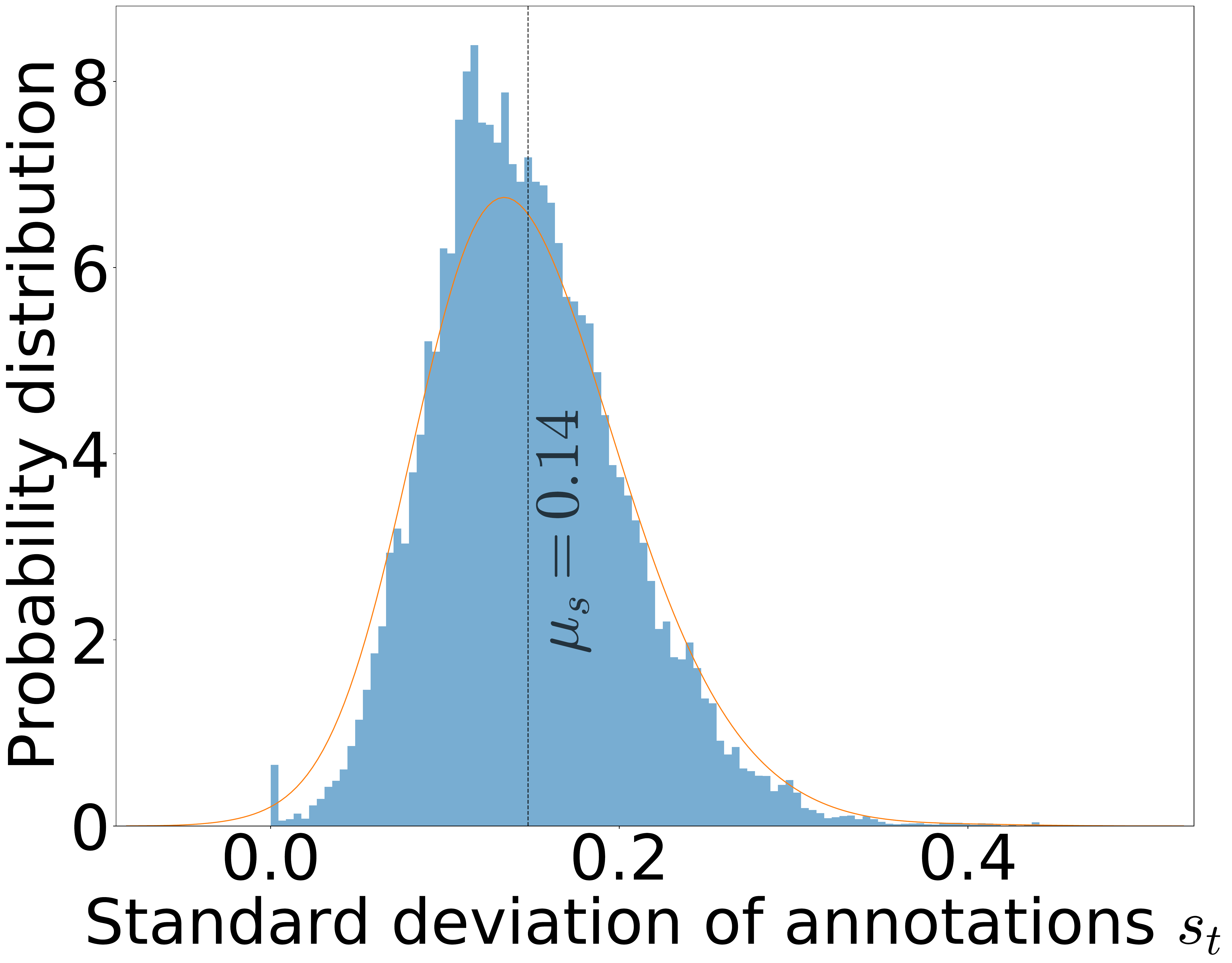}
        \caption{For valence.}
        \captionsetup{justification=centering}
        \label{Fig:s_dist_valence}
     \end{subfigure}
\caption{Distribution of standard deviations $s_t$ in dataset \cite{avec16}}
\label{Fig:s_dist}
\end{figure}


To validate our model, we use the AVEC'16 \cite{avec16} version of the RECOLA dataset \cite{recolaDB}. In this work, we only utilize the audio signals collected at 16 kHz, from the multimodal signals recorded. The dataset consists of continuous arousal and valence annotations by $a=6$ annotators at $40$~ms frame-rate. As illustrated in Figure \ref{Fig:s_dist}, in the AVEC'16 \cite{avec16} dataset, arousal and valence annotations are distributed on average with $\mu_{{m}} = 0.01$ and $\mu_{{m}} = 0.11$, and $\mu_{s} = 0.23$ and $\mu_{s} = 0.14$, respectively, where $\mu_{s} = \dfrac{1}{T} \sum_{t=1}^{T}s_t$. This reveals the significant level of subjectivity present in the dataset, where $s_t$ distributions are heavy-tailed with usually high $s_t$ and $\mu_{s}$. The dataset is divided into speaker disjoint partitions for training, development and testing, with nine $300$~s recordings each. As the annotations for the test partition are not publicly available, all results are computed on the development partition.

\subsection{Baselines}

To evaluate the performance of our proposed approach, we use MTL- and BBB-based uncertainty models \cite{Prabhu2021EndToEndLU}. From \cite{han2020exploring} we use the perception uncertainty (\emph{MTL PU}) and single-task models (\emph{STL}), and from \cite{Prabhu2021EndToEndLU} the model uncertainty (\emph{MU}) and \emph{label uncertainty} (\emph{MU$+$LU}) algorithms. Similar to \cite{Prabhu2021EndToEndLU}, for a fair comparison, we reimplemented the baselines from \cite{han2020exploring}, thereby also enabling us to compare the models in-terms of their $s$ estimates, which were not presented in \cite{han2020exploring}. The method proposed in this work, $t$-distribution based label uncertainty, will be called henceforth as \emph{t-LU}, for convenience.





\subsection{Choice of hyperparameters}
The hyperparameters of the \emph{feature extractor} are fixed as suggested in \cite{Tzirakis2018-speech}, similar to \cite{tzirakis2021-mm, tzirakis2021-semspeech}. The hyperparameters of the \emph{uncertainty layer} are adopted from \cite{Prabhu2021EndToEndLU}, who show state-of-the-art results in label uncertainty modeling. For example, as \emph{prior distribution} $P(w)$ a simple Gaussian prior with unit standard deviation $\mathcal{N}(0, 1)$ was used. Similarly, the \emph{posterior distribution} $P(w|D)$ is initialized using the same heuristics. 


In this work, we assume a Gaussian on $\widehat{\mathcal{Y}}_t$, and noted previously that $n \ge 30$ is required for the assumption to hold. In this light, and keeping the time-complexity in mind, we fixed $n = 30$. For training, we use the Adam optimizer with learning rate $10^{-4}$. The batch size used was 5, with a sequence length of 300 frames, $40$~ms each. All the models were trained for a fixed 100 epochs. The best model is selected and used for testing when best $\mathcal{L}$ \eqref{eq:end-to-end_loss} is observed on train partition.


\begin{table*}[t!]
\centering
\caption{Comparison on mean $m$, standard deviation $s$, and label distribution estimations $\mathcal{Y}$, in terms of $\mathcal{L}_{\text{ccc}}(m)$, $\mathcal{L}_{\text{ccc}}(s)$, and $\mathcal{L}_\text{KL}$, respectively. Larger CCC indicates improved performance as indicated by $\uparrow$. Lower KL indicates improved performance as indicated by $\downarrow$. ** indicates that the respective approach achieves statistically significant better results than \emph{all} other approaches in comparison. * indicates that it achieves statistically significant better results over \emph{only some} of the approaches in comparison.}
    \begin{subtable}{.48\textwidth}
        \centering
            \caption{For arousal}
            \begin{tabular}{lccc}
                \hline
                                        
                    & $\mathcal{L}_{\text{ccc}}(m) \uparrow$    & $\mathcal{L}_{\text{ccc}}(s) \uparrow$    & $\mathcal{L}_\text{KL} \downarrow$     \\
                \hline
                
                STL  \cite{han2020exploring}    & 0.7192        & -                 & -               \\ 
                MTL PU  \cite{han2020exploring} & 0.7336        & 0.2861         & 0.7965         \\ 
                MU  \cite{Prabhu2021EndToEndLU}      & 0.7559        & 0.0764          & 0.6900         \\
                MU$+$LU  \cite{Prabhu2021EndToEndLU} & 0.7437        & {0.3402}    & {0.2576}\\
                \textbf{\emph{t}-LU (proposed)} & \textbf{0.7665**} & \textbf{0.3752**} & \textbf{0.2349**}  \\
                \hline
            \end{tabular}
        \label{tab:arousal_quant_results}
    \end{subtable}
    \quad
    \begin{subtable}{.48\textwidth}
        \centering
            \caption{For valence}
            \begin{tabular}{lccc}
                \hline
                                        
                    & $\mathcal{L}_{\text{ccc}}(m) \uparrow$    & $\mathcal{L}_{\text{ccc}}(s) \uparrow$    & $\mathcal{L}_\text{KL} \downarrow$     \\
                \hline

                STL  \cite{han2020exploring}    & 0.3878               & -                 & -                   \\ 
                MTL PU  \cite{han2020exploring} & \textbf{0.4163}  & 0.0292                & 0.9981                \\ 
                MU  \cite{Prabhu2021EndToEndLU}  & 0.3248           & 0.0359                 & 0.6334               \\
                MU$+$LU  \cite{Prabhu2021EndToEndLU}    & 0.2831           & {0.0422}                 & 0.4054      \\
                \textbf{\emph{t}-LU (proposed)} & {0.3768*}       & \textbf{0.0481*}   & \textbf{0.3914*}     \\
                \hline
            \end{tabular}
        \label{tab:valence_quant_results}
    \end{subtable}
\label{result:quant_results}
\end{table*}

\begin{figure*}[t!]
     \centering
     \begin{subfigure}[b]{0.49\textwidth}
        \includegraphics[width=\textwidth]{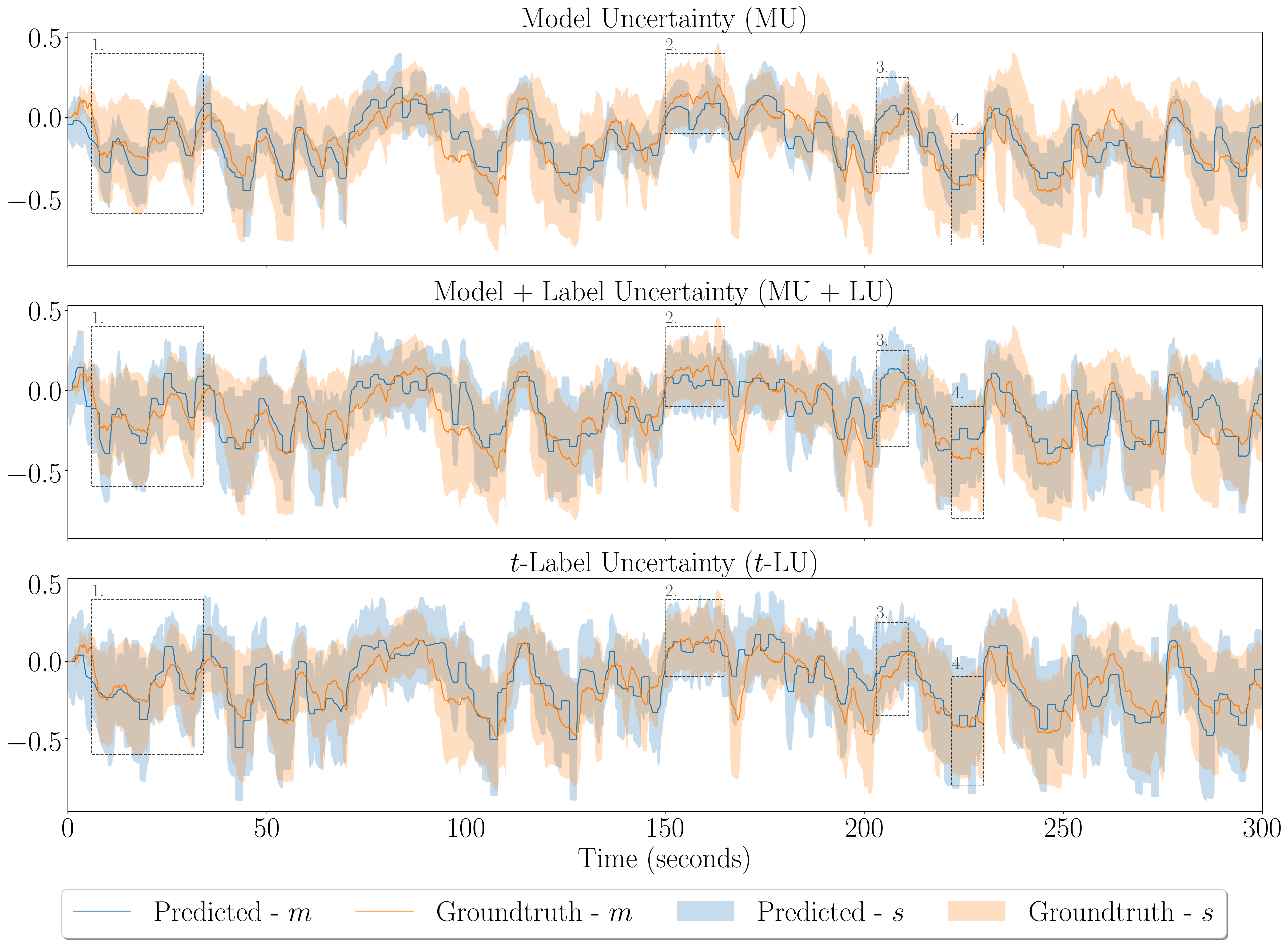}
        \caption{For arousal.}
        \captionsetup{justification=centering}
        \label{Fig:results-pred-arousal}
     \end{subfigure}
     \hfill
     \begin{subfigure}[b]{0.49\textwidth}
        \includegraphics[width=\textwidth]{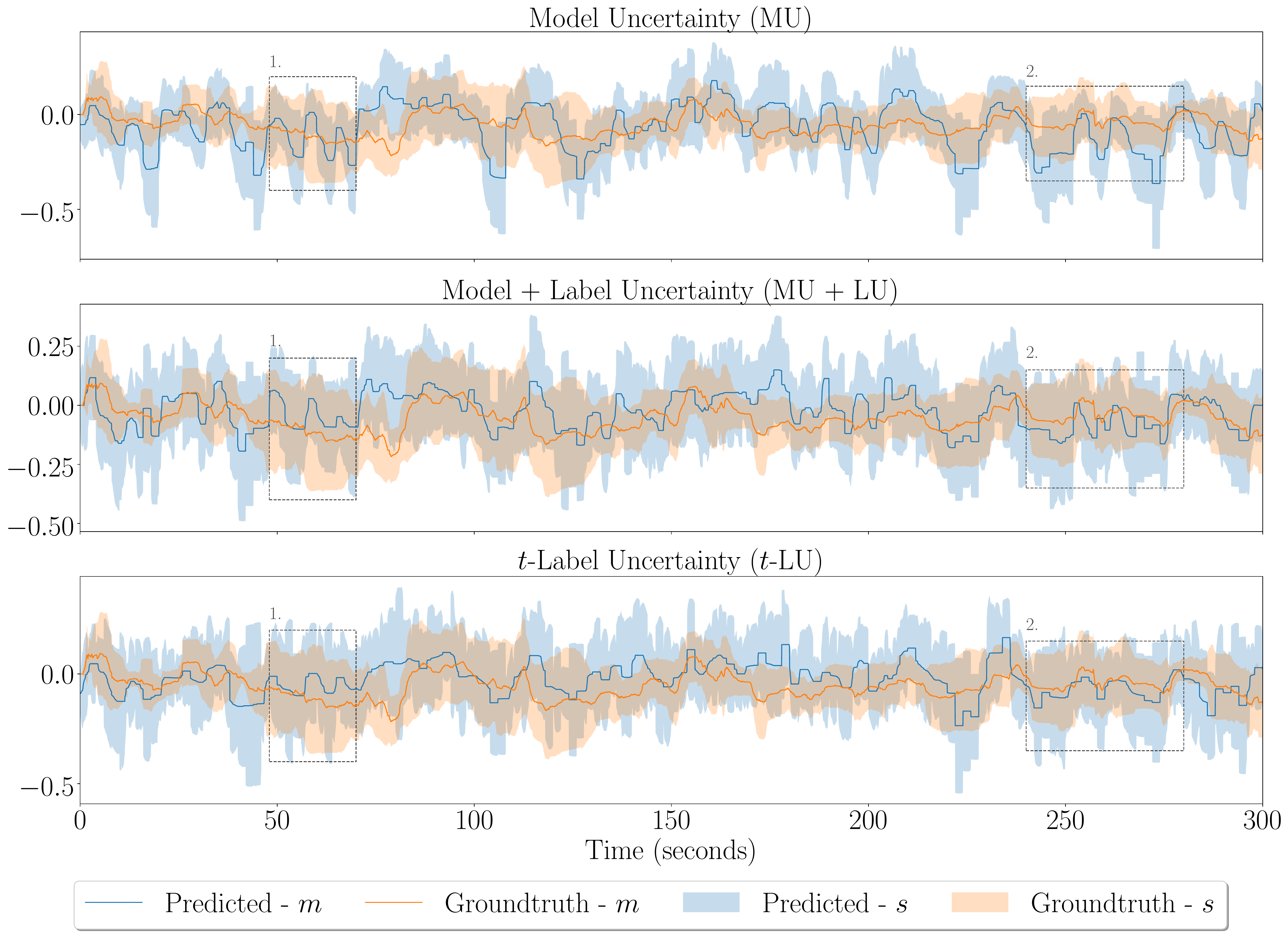}
        \caption{For valence.}
        \captionsetup{justification=centering}
        \label{Fig:results-pred-valence}
     \end{subfigure}
\caption{Label distribution $\mathcal{Y}_t$ estimation  results for a test subject.}
\label{Fig:qual_results}
\end{figure*}

\subsection{Validation measures}
To validate the proposed method's \emph{mean} and \emph{standard deviation} estimates, we use $\mathcal{L}_\text{CCC}(m)$ and $\mathcal{L}_\text{CCC}(s)$ metrics, respectively, widely used in literature \cite{Tzirakis2018-speech, tzirakis2021-semspeech, han2020exploring}. However, $\mathcal{L}_\text{CCC}(m)$ and $\mathcal{L}_\text{CCC}(s)$ validates mean and standard deviation estimates \emph{separately}. To further jointly validate mean and standard deviation estimates, as label distribution $\widehat{\mathcal{Y}}_t$, we use the $\mathcal{L}_\text{KL}$ measure. A similar measure is used in \cite{Prabhu2021EndToEndLU}, but with a Gaussian assumption on $\mathcal{Y}_t$ and hence can be biased. However, for a fair comparison, we validate all the models in comparison using $\mathcal{L}_\text{KL}$ based on their respective distribution assumptions on $\mathcal{Y}_t$, as the models are trained in a similar fashion. The proposed \emph{t-LU} method is validated and trained on the $t$-distribution $\mathcal{L}_\text{KL}$ \eqref{loss:final-KL}, and the baselines from \cite{han2020exploring} and \cite{Prabhu2021EndToEndLU} are validated and trained on Gaussian $\mathcal{L}_\text{KL}$ \eqref{eq:Gauss-KL}. Nevertheless, from the experiments we also noted that the proposed \emph{t-LU} performs better in-terms of both \eqref{loss:final-KL} and \eqref{eq:Gauss-KL}.







\section{Results and Discussion}

Table \ref{result:quant_results} shows the average performance of the baselines and the proposed model, in terms of their mean $m$, standard deviation $s$, and distribution $\widehat{\mathcal{Y}}_t$ estimations, $\mathcal{L}_{\text{CCC}}(m)$, $\mathcal{L}_{\text{CCC}}(s)$ and $\mathcal{L}_{KL}$, respectively. Comparisons with respect to $\mathcal{L}_{\text{CCC}}(s)$ and $\mathcal{L}_{KL}$ are not presented for the STL as this algorithm does not contain uncertainty modeling. Statistical significance is estimated using one-tailed $t$-test on error distributions, asserting significance for \emph{p}-values $\leq0.05$, similar to \cite{sridhar2021generative}.

\subsection{Comparison on mean estimates}
In terms of mean estimates $\mathcal{L}_{\text{CCC}}(m)$ of \emph{arousal}, Table \ref{tab:arousal_quant_results} shows that the proposed \emph{t}-LU model performs best in comparison with the baselines, with statistical significance. While the \emph{t}-LU model achieves an $\mathcal{L}_{\text{CCC}}(m)$ of $0.7665$, the BBB-based baselines, MU$+$LU and MU, achieve $0.7437$ and $0.7559$, respectively. This also reveals the superiority of the proposed $t$-distribution $\mathcal{L}_{KL}$ \eqref{loss:final-KL} over the Gaussian $\mathcal{L}_{KL}$ \eqref{eq:Gauss-KL}, with \emph{t}-LU outperforming MU$+$LU. Moreover, in \cite{Prabhu2021EndToEndLU}, it was noted that training on KL loss \emph{with the Gaussian assumption} \eqref{eq:Gauss-KL} makes a compromise on $\mathcal{L}_{\text{CCC}}(m)$ performances with improving $\mathcal{L}_{\text{CCC}}(s)$. However, the proposed \emph{t}-LU is free from this compromise with \emph{t}-LU outperforming MU. Finally, we also note that the proposed \emph{t}-LU performs significantly better than the MTL-based uncertainty baseline MTL PU, and also the STL which does not model uncertainty.



In terms of mean estimates $\mathcal{L}_{\text{CCC}}(m)$ of \emph{valence}, concerning Table \ref{tab:valence_quant_results}, it is noted that the the proposed \emph{t}-LU performs significantly better than the BBB-based models, MU$+$LU and MU. While the \emph{t}-LU model achieves an $\mathcal{L}_{\text{CCC}}(m)$ of $0.3768$, the BBB-based baselines, MU$+$LU and MU, achieve $0.2831$ and $0.3248$, respectively. Similar to arousal, for valence, we see that \emph{t}-LU is free from compromises on $\mathcal{L}_{\text{CCC}}(m)$, as in \cite{Prabhu2021EndToEndLU}. However, the proposed \emph{t}-LU does not improve over the MTL-based baselines, MTL PU and STL. It is a common trend in SER literature that the audio modality inadequately explains the valence dimension of emotion \cite{tzirakis2021-mm}. However, a probable explanation for this is that the MTL-based architectures are generally better than the BBB-based, in-terms of $\mathcal{L}_{\text{CCC}}(m)$ of valence. Results present by Han et al. \cite{han2020exploring} also show similar behaviour where MTL-based architectures show significant improvements \emph{specifically in mean estimates of valence} \cite{han2020exploring}.




\subsection{Comparison on uncertainty estimates}

While the proposed \emph{t}-LU achieves significantly improved results over the baselines for mean estimates, especially for the arousal dimension, the main goal of this paper is to aptly capture label uncertainty in emotions, in terms of $\mathcal{L}_{\text{CCC}}(s)$ and $\mathcal{L}_{KL}$. Concerning the uncertainty estimates in arousal, Table \ref{tab:arousal_quant_results} shows that the proposed \emph{t}-LU achieves state-of-the-art results, with significant improvements over the baselines. For instance, the MU$+$LU model, the best performing baseline, achieves a $\mathcal{L}_{\text{CCC}}(s)$ and $\mathcal{L}_{KL}$ of $0.3402$ and $0.2576$, respectively, while \emph{t}-LU significantly improves by achieving $0.3752$ and $0.2349$, respectively. This performance, in terms of both the measures, explains the advantage of the $t$-distribution based KL loss term \eqref{loss:final-KL} in label uncertainty modeling. The $t$-distribution $\mathcal{L}_{KL}$, as seen in Figure \ref{fig:kl-analysis}, promotes the model to fit on a more relaxed $s_t$ and penalizes more for tighter standard deviations, thereby leading to better capturing the label distribution.



For valence, unlike the $\mathcal{L}_{\text{CCC}}(m)$ performance,  Table \ref{tab:valence_quant_results} shows that the proposed \emph{t}-LU achieves improved performance in-terms of the \emph{uncertainty estimates}, over all the baselines in comparison. It is further noted that \emph{t}-LU improves with statistical significance over the MTL-based baselines, however improves without statistical significance in comparison with the BBB-based baselines. While the MU$+$LU, best performing baseline, achieves a $\mathcal{L}_{\text{CCC}}(s)$ and $\mathcal{L}_{KL}$ of $0.0422$ and $0.405$, respectively, \emph{t}-LU improves by achieving $0.0481$ and $0.3914$, respectively. However, the lack of statistical significance over some baselines, as well as the generally rather low performance of all approaches, could be owed to the common observation that speech inadequately explains valence \cite{tzirakis2021-mm}.

\subsection{Qualitative analysis on distribution estimation}
To further validate the results, we plot the mean $m$ and standard deviation $s$ of the estimated distributions for a test subject, seen for arousal in Figure \ref{Fig:results-pred-arousal}, and valence in Figure \ref{Fig:results-pred-valence}. Moreover, some parts of the plots are boxed and numbered to note performance differences. For arousal, in Figure \ref{Fig:results-pred-arousal}, further backing the results in Table \ref{tab:arousal_quant_results}, the proposed $t$-LU model better captures $m$ and $s$ of the annotation distribution, in comparison with MU and MU$+$LU. For example, in the box numbered \emph{2} in Figure \ref{Fig:results-pred-arousal}, the proposed $t$-LU captures the whole annotation distribution $\mathcal{Y}_t$, resembles the \emph{Ground-truth - s} best. This further highlights the robustness of training on a relaxed $\sigma_t$ through a $t$-distribution. Moreover, in box \emph{2}, we also note that $t$-LU, by best capturing $s$, also improves notably in terms of the mean estimates $m$. This can be noted in all boxes \emph{1}-\emph{4}, where $t$-LU best captures the whole arousal annotation distribution, by improving on both $s$ and $m$ estimations. 

For valence, in Figure \ref{Fig:results-pred-valence}, backing results in Table \ref{tab:valence_quant_results}, we note that the proposed $t$-LU improves significantly in-terms of the mean estimates $m$, with only small improvements on standard deviation estimates $s$. This can be seen for instance in box \emph{1}, where $t$-LU notably improves in terms of $m$ estimations, over MU and MU$+$LU, while only small improvements in terms of $s$ estimations can be observed. Overall, we note that the proposed speech-based uncertainty model better captures the annotation distribution of arousal than that of valence.
 
\subsection{Analysis on training loss curve}
\begin{figure}[t!]
     \centering
     \begin{subfigure}[b]{0.24\textwidth}
        \includegraphics[width=\textwidth]{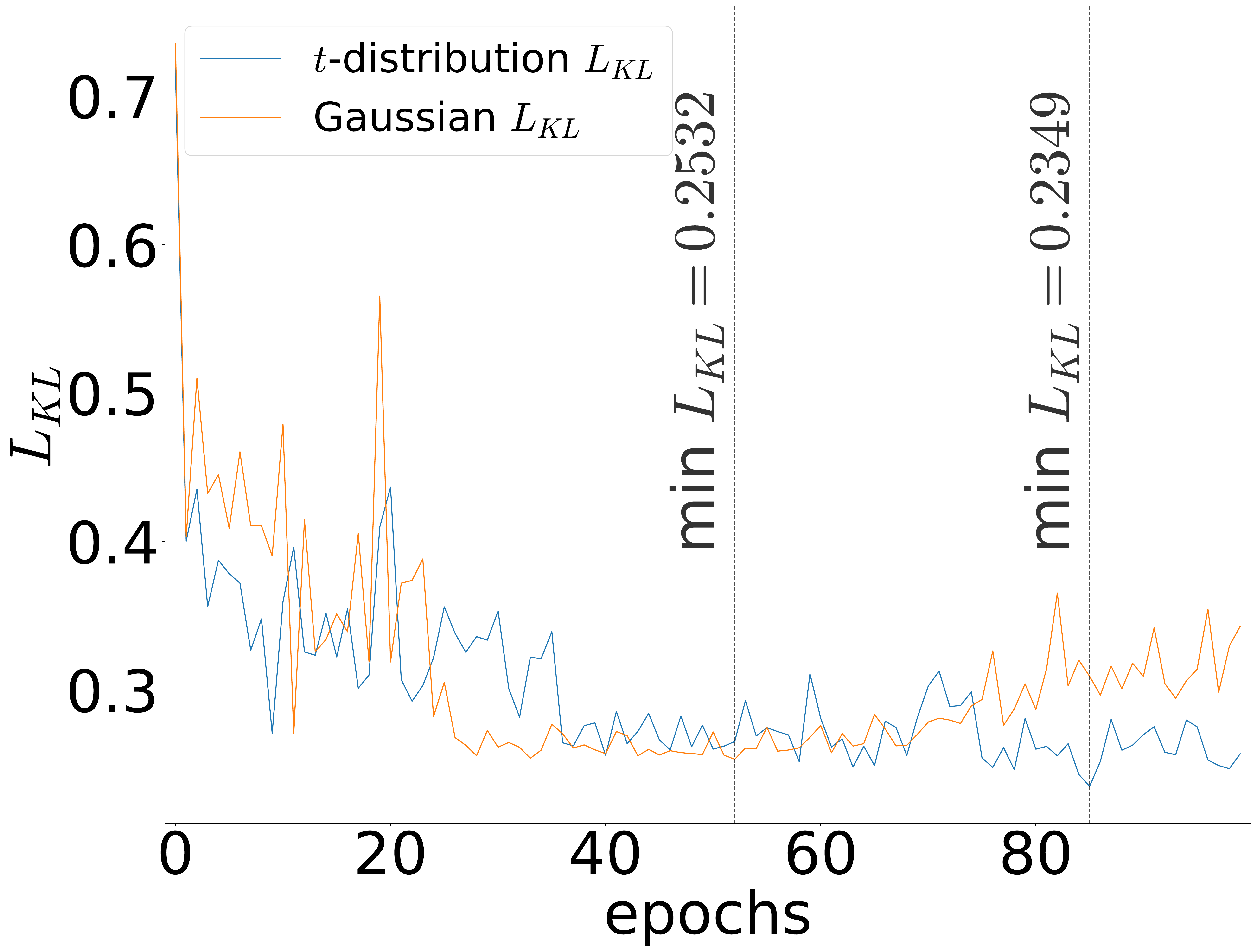}
        \caption{For arousal.}
        \captionsetup{justification=centering}
        \label{Fig:loss_curves_arousal}
     \end{subfigure}
     \hfill
     \begin{subfigure}[b]{0.24\textwidth}
        \includegraphics[width=\textwidth]{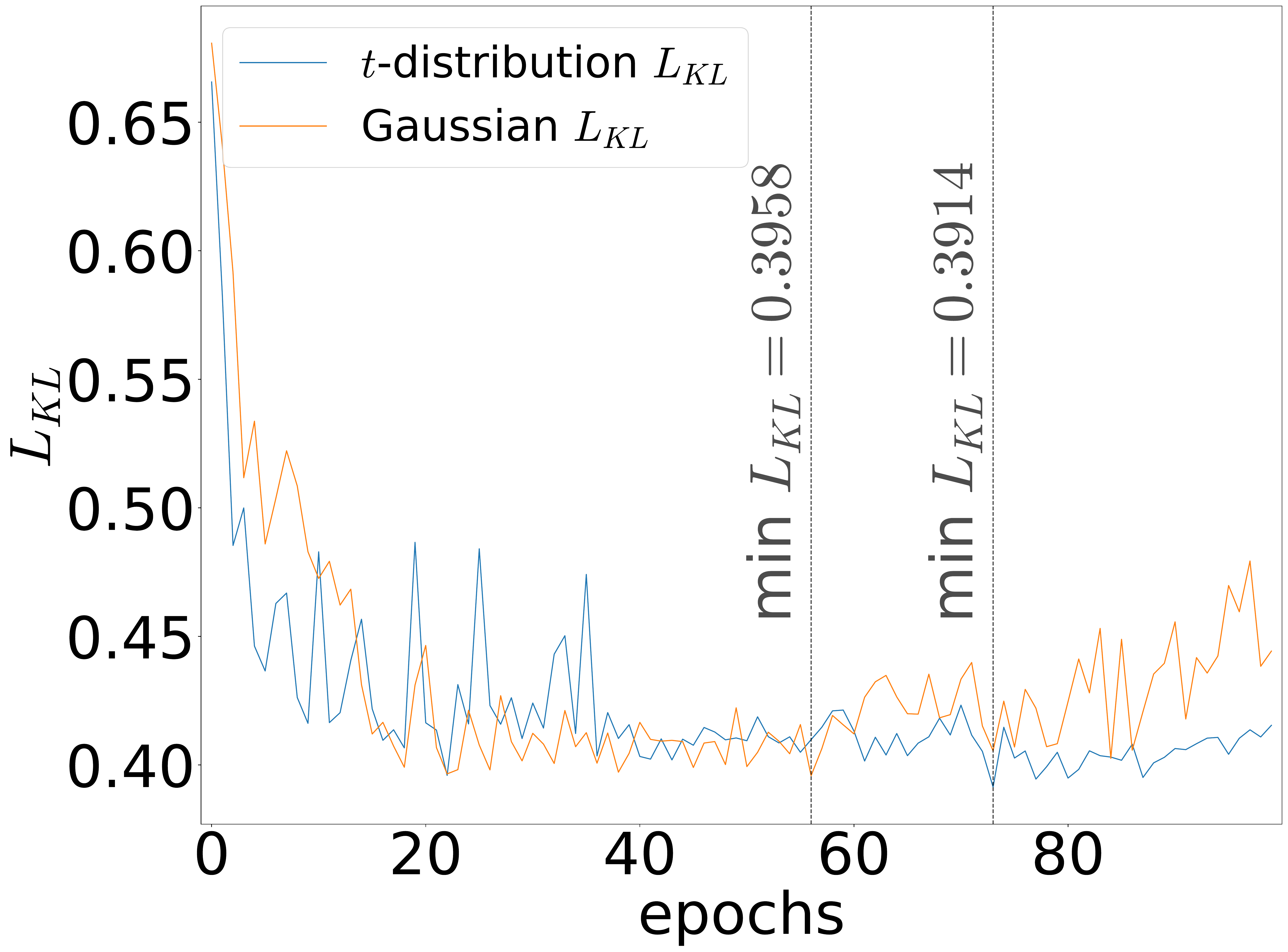}
        \caption{For valence.}
        \captionsetup{justification=centering}
        \label{Fig:loss_curves_valence}
     \end{subfigure}
\caption{Loss curve comparison between Gaussian $\mathcal{L}_{KL}$ \eqref{eq:Gauss-KL} and proposed $t$-distribution $\mathcal{L}_{KL}$ \eqref{loss:final-KL}.}
\label{Fig:loss_curves}
\end{figure}


To further study the advantages of the proposed $t$-distribution $\mathcal{L}_{KL}$ \eqref{loss:final-KL} during the training phase, we compare the testing loss curve of \eqref{loss:final-KL} with the Gaussian $\mathcal{L}_{KL}$ in MU$+$LU \eqref{eq:Gauss-KL}. The comparison can be seen for arousal in Figure \ref{Fig:loss_curves_arousal}, and for valence in Figure \ref{Fig:loss_curves_valence}. 

For both the arousal and the valence dimension, Figure \ref{Fig:loss_curves} illustrates two crucial advantages of the proposed $t$-distribution $\mathcal{L}_{KL}$ \eqref{loss:final-KL} as a loss term in the training phase. Firstly, we see that in the initial epochs, before epoch 20, the proposed loss term converges quicker than the Gaussian $\mathcal{L}_{KL}$ \eqref{eq:Gauss-KL}. This is the result of the proposed $\mathcal{L}_{KL}$ \eqref{loss:final-KL} loss term which penalizes more for lower $s_t$ values, in comparison to the Gaussian $\mathcal{L}_{KL}$ \eqref{eq:Gauss-KL}, as seen in Section \ref{Section:kl-analysis}, thereby achieving faster convergence. Secondly, it is noted that during the later epochs, after 60 epoch, the Gaussian $\mathcal{L}_{KL}$ \eqref{eq:Gauss-KL} shows signs of overfitting with increasing testing loss. However, at the same time, the proposed $\mathcal{L}_{KL}$ \eqref{loss:final-KL} converges to the best minima during the later epochs. The proposed $\mathcal{L}_{KL}$ achieves minima $\mathcal{L}_{KL}$ at the epoch 85, with $\mathcal{L}_{KL}$ of 0.2349 for arousal and 0.3914 for valence, while the Gaussian achieves a minima well before the later epochs, at epoch 54, with $\mathcal{L}_{KL}$ of 0.2532 for arousal and 0.3958 for valence. The proposed $\mathcal{L}_{KL}$ is free from overfitting in the later stages of training and also learns the optima at this stage. This behaviour can be attributed to the nature of the proposed $\mathcal{L}_{KL}$ which promotes the model to learn a more relaxed $s_t$, as seen in Section \ref{Section:kl-analysis}, thereby introducing more regularization to the model, resulting in preventing overfitting and converging on an improved $s_t$.

\section{Conclusions}
Label uncertainty modeling in emotion recognition is commonly approached by assuming a Gaussian distribution on the ground-truth emotion annotations, however with only limited annotations. In contrast, in this work, we assumed a Student's $t$-distribution on the ground-truth emotion annotations, which also accounts for the number of annotations available. This is the first time in literature an attempt was made to handle the limited emotion annotations available, from a machine learning perspective. For this, we proposed and derived a KL divergence based loss term that aims to capture emotion annotation distribution \emph{as a $t$-distribution}. The derived $t$-distribution loss term is also mathematically more sound than the Gaussian assumption, for limited annotations. Subsequently, we showed that the proposed $t$-distribution loss term leads to training on a relaxed standard deviation, which is adaptable with respect to the number of annotations available. Moreover, we also validated our approach on a publicly available dataset. Quantitative analysis of the results showed that the proposed $t$-distribution loss term improves over the Gaussian assumption with state-of-the-art results in mean and standard deviation estimations, in-terms of both the CCC and KL divergence measures. Finally, we also showed, through the analysis of the loss curves, that the proposed loss term leads to faster and improved convergence, and is less prone to overfitting. 

\bibliographystyle{IEEEtran}
\bibliography{main}

\end{document}